\documentclass[aps,twocolumn,showpacs,showkeys,amsmath,amssymb]{revtex4}
\usepackage[dvips]{graphicx}
\usepackage{dcolumn}
\usepackage{bm}

\begin{document}

\title{Role of the E$_{2g}$ phonon in the superconductivity of MgB$_{2}$: a Raman
scattering study}

\author{H. Martinho}
\email{hercules@ifi.unicamp.br}
\homepage{http://www.ifi.unicamp.br/gpoms}

\author{C. Rettori}

\author{P. G. Pagliuso}

\affiliation{Instituto de F\'{\i}sica ``Gleb Wataghin'' UNICAMP,
13083-970 Campinas, SP, Brazil.\\}

\author{A. A. Martin}
\affiliation{Instituto de Pesquisa e Desenvolvimento - UNIVAP,
12.244-050, S\~{a}o Jos\'{e} dos Campos, SP, Brazil.\\}

\author{N. O. Moreno}

\author{J. L. Sarrao}

\affiliation{Los Alamos National Laboratory, Los Alamos, New
Mexico 87545, U.S.A.\\}

\begin{abstract}
Temperature-dependent Raman scattering studies in polycrystalline
MgB$_{2}$\ ($10<T<300$\ K) reveal that the E$_{2g}$ phonon does
not experience any self-energy renormalization effect across the
superconducting critical temperature $T_{C}$\ $\approx 39$\ K, in
contrast with most of the current theoretical models. \ In the
presence of our results, those models must be reviewed. The
analysis of the temperature dependence of the E$_{2g}$ phonon
frequency yields an isobaric Gr\"{u}neisen parameter of $\mid
\gamma _{E_{2g}}\mid \lesssim 1$, smaller than the value of $3.9$
obtained from isothermal Raman experiments under pressure. It is
suggested that this apparent disagreement can be explained in
terms of pressure-induced changes of the topology of the Fermi
surface.

\end{abstract}



\maketitle

MgB$_{2}$ has attracted much recent interest due to its
remarkable physical properties such as: i) relatively high
superconducting transition, $T_{C}$ $=$ $39$ K, for a binary
compound with a simple crystal structure, ii) large and
anisotropic coherence lengths, critical fields and current
densities, and iii) critical currents that are not limited by
grain boundaries (absence of weak link effects).\cite{Buzea}

MgB$_{2}$\ forms in a hexagonal structure with space group P6/mmm
(D$_{6h}^{1}$). The B atoms are located on a primitive honeycomb
lattice consisting of graphite-type sheets. The B$_{2}$-layers
are intercalated with Mg-layers that also form a honeycomb
lattice with a Mg atom in the center. For this space group,
factor-group analysis predicts four modes at the $\Gamma $ point:
E$_{u}+$ A$_{2u}+$ E$_{2g}$ $+$ B$_{1g}$, where only the E$_{2g}$
mode\ is Raman-active and the B$_{1g}$ mode is silent. The
E$_{2g}$ phonon is a doubly degenerate in-plane B-B
bond-stretching mode\cite{Liu} with nonvanishing Raman tensor
elements $(\alpha _{xx}-\alpha _{yy})$ and $\alpha _{xy}$. First
principles lattice dynamics calculations indicate that these
modes would be observed at $327$ cm$^{-1}$(E$_{u}$)$,$ $405$ cm$^{-1}$(A$%
_{2u}$), $572$ cm$^{-1}$(E$_{2g}$) and $702$ cm$^{-1}$(B$_{1g}$). \cite
{Bohnen}

In analogy with high $T_{C}$ superconductors (HTS), Hall effect
measurements \cite{Hall effect} indicate that the charge carriers
in MgB$_{2}$ are holes with a hole density at $300$ K of
$1.7-2.8\times $10$^{23}$ holes/cm$^{3}$.\ In fact, hole-mediated
superconductivity has been proposed by An and Pickett for
MgB$_{2}$.\cite{An} These authors attribute the relatively high
value of
$T_{C}$ to the strong coupling between holes and the in-plane boron phonon, E%
$_{2g}$ modes. According to this model, the holes originate in
the $\sigma $ (sp$^{2}$ orbitals) bands due to charge transfer
from the $\sigma $ bands to the $\pi $ (p$_{z}$ orbitals) bands.
Based on this model, several papers have discussed the possibility
of E$_{2g}$\ phonon being a frozen-in mode, strongly coupled to
the $\sigma $ electronic bands near the Fermi level.\cite
{Liu,Bohnen,Yildirim,Kong} It is claimed that the E$_{2g}$ phonon,
due to its rather strong coupling to the $\sigma $\ electronic
bands, would be highly anharmonic, presenting a very large
linewidth. However, Boeri et al \cite
{Boeri} pointed out that neither the presence at the Fermi level of the $%
\sigma $ bands nor their strong coupling to the E$_{2g}$ phonon
are sufficient to induce such anharmonic effects, and therefore,
the strong anharmonicity would be mainly related to the small
value of the $\sigma $ conduction holes Fermi energy ($\simeq
0.45$ eV) in the unperturbed crystal. Boeri et al also predicted
similar results for heavily hole-doped graphite.

On the other hand, in a perfect harmonic crystal, the equilibrium
size\ would not depend on temperature. Consequently, the phonon
frequency would be temperature independent and its linewidth
should tend to zero. Hence, it is expected that, for strong
anharmonicity, the E$_{2g}$ mode in MgB$_{2}$ would be very broad
and its frequency change strongly with temperature. Isothermal
pressure-dependent Raman scattering results and lattice parameter
measurements by Goncharov et al \cite{Goncharov} have given some
evidence for this anharmonic behavior. They have observed at room
temperature a very broad ($300$ cm$^{-1}$) Raman mode at $620$
cm$^{-1}$ associated to the E$_{2g}$ mode and an anomalously
large mode Gr\"{u}neisen parameter, $\gamma _{E_{2g}}=3.9\pm 0.4$.

The correlation between $T_{C}$ and the resistivity ratio between
room and near-$T_{C}$ temperatures, known as the Testardi
correlation, is an evidence in favor of a dominant electron-phonon
mechanism in MgB$_{2}$.\cite{Buzea} Moreover, the boron isotope
effect shows that the boron modes may play a significant role in
the MgB$_{2}$ superconductivity. \cite{Bisotope} Besides, $T_{C}$
$=$ $39$ K for MgB$_{2}$ is at the extreme end of $T_{C}$'s
predicted by the BCS theory\cite{BCStc} that would require an
electron-phonon coupling (EPC) of $\lambda \approx 1$. \cite{Liu}
However, experimental results\cite{EPCexp} give $\lambda \approx
0.6-0.7$ while first principle
calculations\cite{Bohnen,Kong,Kortus} yield an EPC of $\lambda
\approx 0.7-0.9$. Nevertheless, in these papers there is a
consensus about the importance of the E$_{2g}$ phonon in the
superconducting mechanism of\ MgB$_{2}$. Particularly, Liu et al
\cite{Liu} suggested that the strongest EPC arises along the
$\Gamma -A$ line for the E$_{2g}$ mode, and a $12\%$ hardening
below $T_{C}$ at the $\Gamma $ point is predicted. Therefore, the
aim and the important contribution of the present work is to
study, by means of Raman scattering, the temperature dependence of
the E$_{2g} $\ mode both below and above $T_{C}$. Although other
groups had performed temperature dependence studies of the Raman
spectra of MgB$_{2}$, e.g. superconducting gap studies by Raman
scattering in refs. \cite{Gap1,Gap2}, we present the first
systematic investigation of the temperature behavior of the E2g
phonon in this compound addressing its possible role in the
superconducting mechanisms of MgB2.

Raman spectroscopy is an excellent technique to investigate
low-energy elementary excitations, particularly in
superconductors. It was one of the first spectroscopic techniques
to reveal both the existence of the superconducting gap and its
strong coupling to some of the active Raman
phonons.\cite{CardonaRev} In general, for strong electron-phonon
coupling, the theoretical models predict that a phonon with the
symmetry of the gap and comparable energy should soften or harden
below $T_{C}$, depending on whether the phonon frequency is
higher or lower than twice the superconducting energy gap,
$2\Delta _{0}$.\cite{Zeyher,Devereaux} Also, the phonon linewidth
should change in the superconducting state; phonons with
energy below $2\Delta _{0}$ should sharpen and phonons with energy above $%
2\Delta _{0}$ should broaden.\cite{Zeyher,Devereaux} These behaviors have
been observed in most of the HTS.\cite{Friedl,Leach,AAM} Besides, the
temperature dependence of the phonon frequency and linewidth measured by
Raman scattering can give some insight about the anharmonicity of the E$%
_{2g} $ mode. In general, due to the lattice expansion
contribution, \cite {Menendez} the harmonic phonon frequency is
temperature-dependent and in the lowest order is given by:
\begin{equation}
\omega \left( T\right) =\frac{\omega _{0}}{2}\left( 1+e^{-3\gamma
\int\limits_{0}^{T}\alpha (T^{\prime })dT\prime }\right)  \eqnum{1}
\label{eq.(1)}
\end{equation}
where $\gamma $ is the Gr\"{u}neisen parameter, $\alpha $ is the
coefficient of thermal expansion, and $\omega _{0}\equiv \omega
\left( T\rightarrow 0\right) $. The linewidth also has a
temperature dependence that in the lowest order is given
by\cite{Menendez}
\begin{equation}
\Gamma \left( \omega _{0},T\right) =\Gamma \left( \omega _{0},0\right) \left[
1+2n\left( \frac{\omega _{0}}{2},T\right) \right]  \eqnum{2}  \label{eq.(2)}
\end{equation}
where $\Gamma \left( \omega ,0\right) $ is the residual linewidth at $%
T\rightarrow 0$ K and $n\left( \omega _{0},T\right) =\left( e^{\frac{\hbar
\omega _{0}}{kT}}-1\right) ^{-1}$ the Bose-Einstein factor. Equation 2
represents the decay of the phonon with frequency $\omega _{0}$ into two
others phonons with half frequencies, $\omega _{0}/2$, and opposite wave
vectors.

Our Raman measurements were made on a polycrystalline MgB$_{2}$ sample
prepared in sealed Ta tubes as described previously \cite{Canfield}. X-ray
powder-diffraction analysis confirmed single-phase purity and an AlB$_{2}$%
-type structure for our MgB$_{2}$ samples. The transition
temperature determined by measuring the real and imaginary
components of \ ac-susceptibility was found to be
$T_{C}\thickapprox $ $39$ K, as indicated in Figure 1. The
collected Raman spectra were dispersed by a Jobin-Yvon (T64000)
spectrograph and detected with a charge-coupled device (CCD)
camera. The scattered signal was polarized with the electric
field direction perpendicular to the grating grooves in order to
maximize the spectrometer response. The $514.5$ nm line of an
Ar$^{+}$ laser was used as the excitation source. The temperature
dependence of the E$_{2g}$ phonon was measured in a pellet of
pressed powder using a laser power\ of $10$ mW on a spot of $\sim
$ $100$ $\mu $m in diameter. The polycrystalline sample was
attached to the cold finger of a temperature controller using a
closed-cycle helium cryostat and measured in a
near-backscattering configuration.\ For the laser power used in
this experiment we estimate a maximum temperature difference
between the sample and thermometer to be $\lesssim 2$ K.

Figure 2 presents few spectra between $10-300$ K. Using a
lorentzian lineshape and a linear background, the frequency and
linewidth of the E$_{2g}$ phonon were extracted and their
temperature dependence shown in Figures 3a and 3b, respectively.
In contrast with some HTS\cite{Misochkoeoutros}\ the integrated
intensity of this peak does not depend on temperature within the
accuracy of our measurements. As mentioned above, among the
active phonons the E$_{2g}$ mode is considered the main candidate
to dominate the electron-phonon coupling.
\cite{Liu,Bohnen,An,Kortus} The large EPC between
the E$_{2g}$\ phonon and the electronic band in the $\Gamma -A$\ line{\em \ }%
led{\em \ }Liu et al\cite{Liu} to predict a $12\%$ hardening, $\Delta \omega
\approx 76$ cm$^{-1}$, of the phonon frequency below $T_{C}$, which should
be easily observable in our Raman experiments. Figure 3a shows that, within
the accuracy of our measurements, the phonon frequency does not present any
change between the normal and superconducting state, i.e., neither softening
nor hardening occurs at\ $T_{C}$\ and between\ $T_{C}$\ and room temperature
a small hardening of \ $\simeq 2$\ cm$^{-1}$\ can be observed. Besides,
Figure 3b shows that the linewidth presents the expected two-phonon
anharmonic decay behavior indicated by the solid line (see discussion below)
and no anomalous temperature dependence of the linewidth is observed near $%
T_{C}$ for this phonon. These results are somewhat surprising in
view of several theoretical
predictions\cite{Liu,Bohnen,An,Kortus} and what has been observed
in most of the HTS.\cite{Friedl,Leach,AAM}

As pointed out by several authors, the linewidth of this mode is
extremely large, indicating strong anharmonicity for this mode.
Particularly, Goncharov et al. \cite{Goncharov} have shown that
the Gr\"{u}neisen parameter related to the E$_{2g}$ mode is very
large, $\gamma _{E_{2g}}=3.9\pm 0.4$. Taking into consideration
the harmonic frequency $\omega _{0}=631$ cm$^{-1}$, a frequency
error bar of $\pm 2$ cm$^{-1}$for the E$_{2g}$ phonon, the $\gamma
_{E_{2g}}$ value obtained by Goncharov, and the thermal expansion
coefficient in the $a$ direction $\alpha _{a}=5.4\times 10^{-6}$
K$^{-1}$ measured by Jorgensen et al, \cite{Jorgensen} the
expected temperature dependence of the E$_{2g}$ frequency can be
calculated. Using $\gamma _{E_{2g}}=3.9\pm 0.4,$\cite{Goncharov}
we obtain temperature dependence of the E$_{2g}$ frequency
illustrated by the area between the two dashed lines in the
Figure 3a. These lines were obtained with eq.(\ref{eq.(1)}) using
the two limiting harmonic frequencies of $633$ cm$^{-1}$ and
$629$ cm$^{-1}$. \ Clearly, the calculated temperature dependence
of the E$_{2g}$\ frequency is not in agrement with the
experimental data. In order to estimate the $\gamma _{E_{2g}}$
value from our data, we present two simulations using
eq.(\ref{eq.(1)}) for two different values of $\gamma _{E_{2g}}$,
$-1.0$ and $1.0$. The area
limited by the two dotted lines corresponds to the expected behavior of the E%
$_{2g}$\ frequency as a function of temperature considering $\gamma
_{E_{2g}}\approx -1.0$. Negative Gr\"{u}neisen parameters are relatively
rare. Some of the modes in negative thermal expansion material ZrW$_{2}$O$%
_{8}$\cite{GruNeg} and in RbI at low temperatures\cite{White} are
examples of compounds, where they have been found.

Finally, the area corresponding to $\gamma _{E_{2g}}\approx 1.0$\
is represented by the area between the two solid lines in the Figure 3a. $%
\gamma _{E_{2g}}\approx 1.0$ is a\ typical value of Gr\"{u}neisen parameter
for \ most of the solids. Then, from the comparison between the data and the
estimations made for different values, we estimate $\mid \gamma
_{E_{2g}}\mid \lesssim 1.0$, in\ disagreement with the large value obtained
by Goncharov et al. According to our result, the strong anharmonicity of the
E$_{2g}$\ mode cannot be attributed to a large Gr\"{u}neisen parameter.

In addition, pressure-induced changes of the topology of the Fermi
surface, related to Lifshitz topological electronic transition,
were invoked \cite{Meletov} to reconcile the smooth pressure
behavior of the $a$, $c$\ parameters and the clear anomalous
E$_{2g}$\ pressure dependence. Those pressure-induced changes
in{\em \ }the Fermi surface may also be the explanation for the
large difference between our $\gamma _{E_{2g}}$\ value and the
one obtained by Goncharov et al. The anharmonic effect on the
linewidth is shown in Figure 3b as a solid line which was obtained
using eq.(\ref{eq.(2)}) with $\omega _{0}=631$ cm$^{-1}$ and
$\Gamma \left( \omega _{0},0\right) =180$ cm$^{-1}$. Notice that
the two-phonon anharmonic decay fits the temperature dependence of
the linewidth data quite well.

In summary, our results indicate that the temperature dependence
of the frequency and linewidth of E$_{2g}$ mode shows no anomaly
at $T_{C},$ revealing that the involvement of the E$_{2g}$
phonon\ near the\ $\Gamma $ point in the electron-phonon
mechanism of superconductivity in MgB$_{2}$ must be revised and
probably other B-modes, far away from the Brillouim zone center,
may be related to the superconducting mechanism in this material.
In analyzing the temperature dependence of the E$_{2g}$
frequency, we estimate a Gr\"{u}neisen parameter of $\mid \gamma
_{E_{2g}}\mid \lesssim 1.0$\ for this mode, which is in
discrepancy with the value estimated from isothermal
pressure-dependent Raman scattering experiments. We suggest that
this apparent disagreement can be accounted for by considering the
pressure-induced changes of the topology of the Fermi
surface.\cite{Meletov}

\section{Acknowledgments}

This work was supported by the Brazilian Agencies CNPq and FAPESP. Work at
LANL was performed under the auspices of the US DOE.

\begin{figure}[bth]
\caption{\label{Fig. 1} Real and imaginary part of
$ac$-susceptibility, $\protect\chi ^{\prime }$ and $\protect\chi
^{\prime \prime }$,taken at $ H_{ac}=1$ Oe and $\protect\nu =1$
kHz. The superconducting critical temperature obtained was
$T_{C}\approx 39$ K.}
\end{figure}

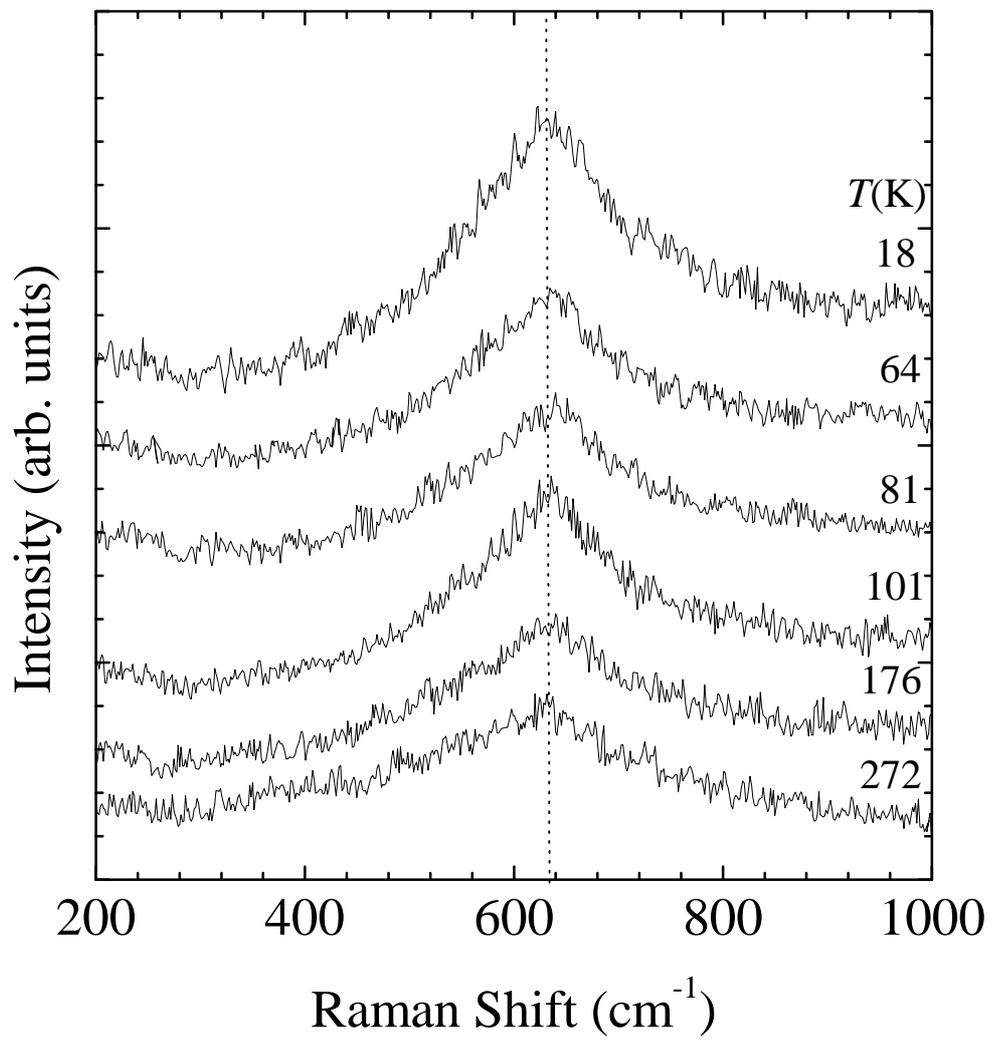
\begin{figure}[bth]
\caption{\label{Fig. 2} Representative of Raman spectra between
$10-300$ K showing the temperature dependence of the E$_{2g}$
mode.}
\end{figure}

\begin{figure}[bth]
\caption{\label{Fig. 3} Temperature dependence of the frequency
(a) and linewidth (b) for the $E_{2g}$ phonon extracted from the
Raman spectra in a polycrystalline sample of MgB$_{2}$. The
solids, dotted, and dashed lines in (a) are simulations using eq.
(1) as explained in the text. The vertical line indicates $T_{C}$
value.}
\end{figure}

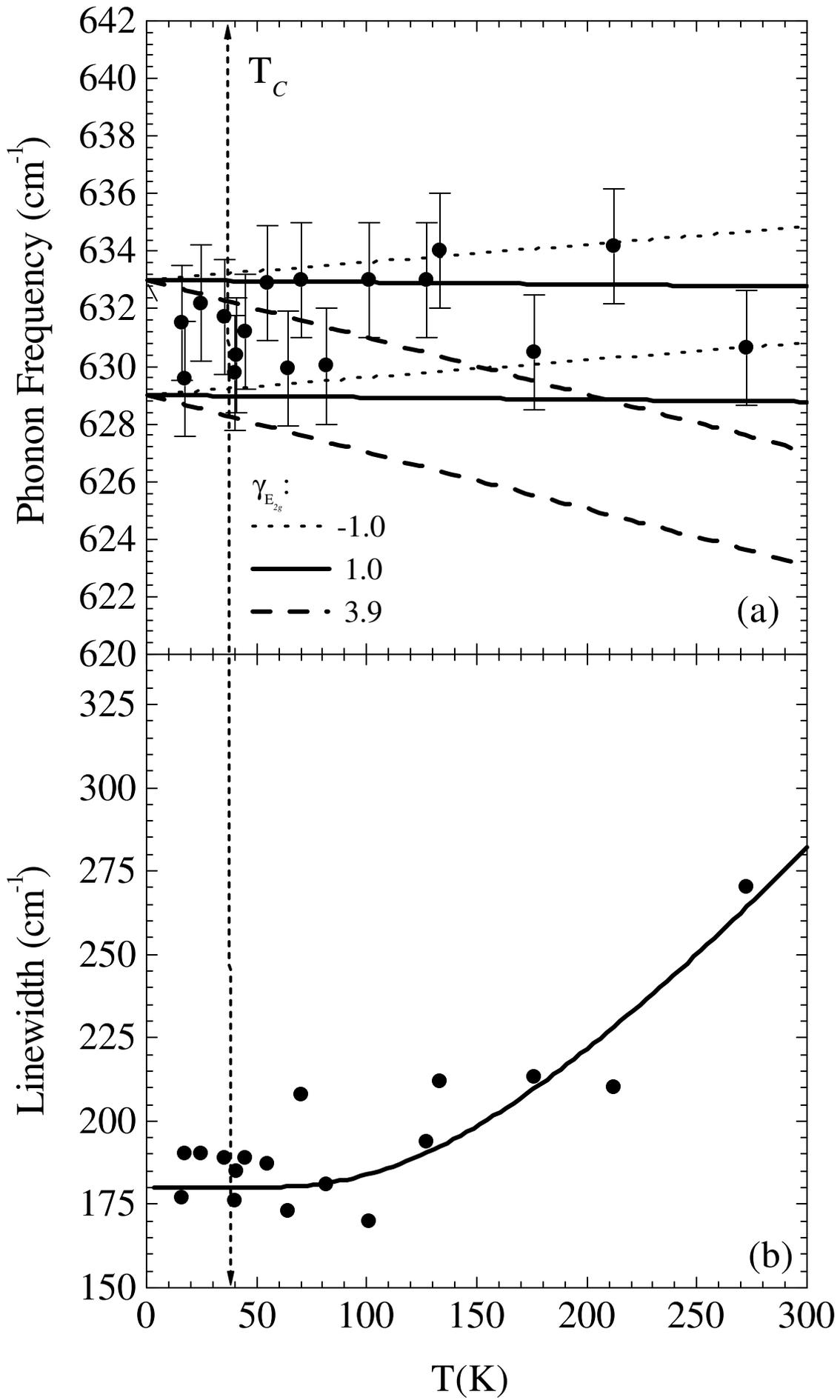
\end{document}